\documentclass{article}
\usepackage{amssymb}

\input{tcilatex}
\begin{document}

\title{\textbf{The problem of time in quantum mechanics}}
\author{M. Bauer \\
Instituto de F\'{\i}sica\\
Universidad Nacional Aut\'{o}noma de M\'{e}xico\\
bauer@fisica.unam.mx}
\maketitle

\begin{abstract}
The problem of time in quantum mechanics concerns the fact that in the Schr%
\"{o}dinger equation time is a parameter, not an operator. Pauli's objection
to a time-energy uncertainty relation analogue to the position-momentum one,
conjectured by Heisenberg early on, seemed to exclude the existence of such
an operator. However Dirac's formulation of electron's relativistic quantum
mechanics (RQM) does allow the introduction of a dynamical time operator
that is self-adjoint. Consequently, it can be considered as the generator of
a unitary transformation of the system, as well as an additional system
observable subject to uncertainty. In the present paper these aspects are
examined within the standard framework of RQM.


\end{abstract}

\section{ Introduction}

In the time dependent Schr\"{o}dinger equation (TDSE) of quantum mechanics
(QM) time appears as a parameter, not as an operator\cite{Pauli,Dirac}.
Pauli's objection to the existence of a time operator, satisfying a
commutation relation $[H,t]=-i\hslash $ as entertained by Heisenberg early
on, is, to quote: "...from the C.R. written above it follows that $H$
possesses continously all eigenvalues from $-\infty $ to $+\infty $, whereas
on the other hand, discrete eigenvalues of $H$ can be present. \textit{We,
therefore, conclude that the introduction of an operator }$\mathit{t}$%
\textit{\ is basically forbidden and the time }$t$\textit{\ must necessarily
be considered as on ordinary number ("c" number) in Quantum Mechanics"}.

Pauli's argument, sustained also by the fact that the system's stability
requires the energy to have a finite minimum, has given rise to a variety of
alternative proposals for a time-energy uncertainty relation and an
extensive discussion of time in quantum mechanics throughout several decades 
\cite{Muga,Muga2,Bauer1,Bush,Briggs,Hilgevoord,Galapon,Boykin}. Within the
time quantities considered one finds parametric (clock) time, tunneling
times, decay times, dwell times, delay times, arrival times or jump times,
i.e., both instantaneous values and intervals. To quote Ref.3:
\textquotedblleft In fact, the standard recipe to link the observables and
the formalism does not seem to apply, at least in an obvious manner, to time
observables\textquotedblright . The extensive experimental confirmation of
the Schr\"{o}dinger equation asserts that this parameter corresponds to the
time coordinate of the laboratory frame of reference, in both QM and RQM
(Relativistic Quantum Mrchanics).

This is the problem of time in quantum mechanics. The questions to be
answered are: 1) Can a time operator be found? 2) What is the status of a
time-energy uncertainty relation? 3) How did the parameter $t$ enter into
the Schr\"{o}dinger equation?

Recently it has been shown that Dirac's formulation of relativistic quantum
mechanics (RQM) does allow the introduction of a dynamical time operator
that is self-adjoint\cite{Bauer}. Consequently, it can be considered as the
generator of a unitary transformation of the system, as well as an
additional system observable subject to uncertainty. In the present paper
these aspects are examined within the standard framework of RQM. The
definition and main properties of the proposed time operator are recalled in
Section 2. Section 3 analyses the effect of the corresponding unitary
transformation. In Section 4 the ensuing time-energy uncertaintinty relation
is derived and shown how it circumvents Pauli's objection. It is also
compared to the Mandelstam-Tamm formulation. Section 5 advances conclusions
and possible developments.

Finally, to explain its presence in the TDSE has led to the consideration of
time as an emergent property arising from the entanglement of a microscopic
system with a classical environment in an overall closed time independent
system, this property being apparent only to an internal observer\cite%
{Briggs2,Moreva}

The proof that the commutation relation $[\hat{x},\hat{p}]=i\hslash $
necessarily implies that the corresponding spectra of $\hat{x}$ and $\hat{p}$
go from $-\infty $ to $+\infty $ continously (as shown in Ref.14) is absent
from most introductory textbooks on quantum mechanics. Also absent is the
fact that the representations of the position and momentum operators in the
configuration and momentum spaces follow from this commutation relation, as
well as the fact that the representations of the state vector in
configuration and momentum spaces are Fourier transform of each other. This
is pedagogically unfortunate, as the student is induced to consider these as
unrelated assumptions. In particular he will miss the connection of the
infinite continuity of these espectra to unitary transformations and group
properties, which is a cornerstone of the further development of quantum
mechanics and of quantum field theory. Appendix A presents a unified
description of the consequences of the commutation relation that could be
incorporated in textbooks..

\section{The dynamical time operator in RQM\protect\cite{Bauer}}

A dynamical self-adjoint "time operator"%
\begin{equation}
\hat{T}=\mathbf{\alpha .\hat{r}/}c\mathbf{+}\beta \tau _{0}
\end{equation}%
has been proposed in analogy to the Dirac free particle Hamiltonian $\hat{H}%
_{D}=c\mathbf{\alpha .\hat{p}}+\beta m_{0}c^{2\text{ }},$ where $\alpha
_{i}(i=1,2,3)$ and $\beta $ are the $4\times 4$ Dirac matrices, satisfying
the anticonmutation relations:%
\begin{equation}
\alpha _{i}\alpha _{j}+\alpha _{j}\alpha _{i}=2\delta _{ij}\ \ \ \ \ \alpha
_{i}\beta +\beta \alpha _{i}=0\ \ \ \ \beta ^{2}=1
\end{equation}
The parameter$\ \tau _{0}$ represents in principle an internal property of
the sysytem, to be determined. In the Heisenberg picture, the time evolution
of the time operator is given by:%
\begin{equation}
\hat{T}(t)=\mathbf{\alpha (}t\mathbf{).\hat{r}(}t\mathbf{)/}c\mathbf{+}\beta
(t)\tau _{0}=\hat{T}(0)+(c\mathbf{\hat{p}}/\hat{H}_{D})^{2}t+oscillating\
terms
\end{equation}%
where use has been made of the following relations\cite%
{Thaller,Greiner,Messiah}:%
\begin{equation}
\mathbf{\alpha (}t\mathbf{)}=\mathbf{\alpha (}0\mathbf{)+\{\alpha (}0\mathbf{%
)-}c\mathbf{\hat{p}}/\hat{H}_{D}\}\{\exp (-2i\hat{H}_{D}t/\hslash )-1\}
\end{equation}%
\begin{equation}
\beta (t)=\beta (0)+\{\beta (0)-m_{0}c^{2}/\hat{H}_{D}\}\{\exp (-2i\hat{H}%
_{D}t/\hslash )-1\}
\end{equation}%
\begin{equation}
\mathbf{\hat{r}(}t\mathbf{)=\hat{r}(}0\mathbf{)+(}c^{2}\mathbf{\hat{p}}/\hat{%
H}_{D})t+i(c\hslash /2\hat{H}_{D})\{\exp (-2i\hat{H}_{D}t/\hslash )-1\}
\end{equation}

\begin{equation}
c\mathbf{\alpha (}0\mathbf{).(}c\mathbf{\hat{p}}/\hat{H}_{D})=\left[ \frac{d%
\mathbf{\hat{r}}}{dt}\right] _{t=0}.\mathbf{(}c\mathbf{\hat{p}}/\hat{H}_{D})=%
\mathbf{(}c\mathbf{\hat{p}}/\hat{H}_{D})^{2}+oscillating\ terms
\end{equation}%
Thus $\hat{T}(t)$\ exhibits a linear dependence on $t$\ with a superimposed
oscillation (Zitterbewegung), as occurs with the time development of the
postion operator $\mathbf{\hat{r}}(t)$.

Following step by step the algebra developed for the Dirac Hamiltonian\cite%
{Thaller,Greiner,Messiah}, the eigenvalue equation%
\begin{equation}
\hat{T}|\tau >=\tau |\tau >
\end{equation}%
yields:%
\begin{equation}
\tau =\pm \ \tau _{r}=\pm \ [(r/c)^{2}+\tau _{0}^{2}]^{1/2}
\end{equation}%
as $\hat{T}^{2}=(\mathbf{\hat{r}}/c)^{2}+\tau _{0}^{2}$. The eigenvalue
spectrum has two continous branches, a positive and a negative one separated
by a $2\tau _{0}$ gap. As $r$ goes from $-\infty $ to $+\infty $\ the
positive branch drops from $+\infty $\ \ to $\tau _{0}$\ when $r=0$ ,\ and
then rises to $+\infty $\ again.\ The negative branch follows the opposite
behaviour.\ 

Each of these eigenvalues is doubly degenerate with respect to the component 
$\boldsymbol{\sigma }$$\boldsymbol{\cdot \hat{r}}/2r$ of the spin along the $%
\boldsymbol{r}$ direction, which commutes with $\hat{T}$. Thus one can find
simultaneous eigenfunctions of $\boldsymbol{\sigma }$$\boldsymbol{\cdot \hat{%
r}}/2r$ and $\hat{T}$, giving rise to altogether four eigenvalue pairs $%
\left\vert \tau ,\sigma \right\rangle $. The "time eigenvectors" are:%
\begin{equation}
|\pm \tau _{r},\pm 1/2>=u_{r}^{i}|\boldsymbol{r}>\ \ \ \ i=1,2,3,4
\end{equation}%
where $|\boldsymbol{r}>$ is the eigenvector of the position operator $%
\boldsymbol{\hat{r}}$ and $u_{r}^{i}$ is a four component spinor independent
of the linear momentum $\boldsymbol{p}$. The four orthonormal spinors $%
u_{r}^{i}$ are listed in Ref.11.

In this formulation, $\tau _{0}$ plays the role of an invariant quantity in
the $(\boldsymbol{r},\tau )$ space, i.e., $\tau _{0}^{2}=\tau ^{2}-(%
\boldsymbol{r}/c)^{2}$, as $m_{0}c^{2}$ plays in the $(\boldsymbol{p},E)$
space, namely $(m_{0}c^{2})^{2}=E^{2}-(c\boldsymbol{p})^{2}$. To mantain the
fundamental indeterminacy modulo $n2\pi $\ ($n$ an integer) in the phase of
the complex eigenfunctions one has to set, for $n=1$\textit{:}%
\begin{equation}
\tau _{0}=2\pi \hbar /<\beta >\varepsilon =h/m_{0}c^{2}
\end{equation}%
This is the de Broglie period\cite{Broglie,Bayliss}. Together with the
Compton wave length, it sets a unified spacetime Compton scale that limits
the wave packets width in space and time before negative energy and negative
time components (particle and antiparticle) occur significantly. Moreover,
it supports the existence of an internal property, the "de Broglie clock"
with a period $\tau _{0}=h/m_{0}c^{2}$\cite{Ferber,Lan,Catillon}.

\section{The time operator as generator of a unitary transformation}

The proposed operator is clearly self-adjoint and therefor can be the
generator of a unitary transformation (Stone's theorem\cite{Jordan}):%
\begin{equation}
\hat{U}_{T}(\varepsilon )=e^{-i\varepsilon \hat{T}/\hslash
}=e^{-i\varepsilon \{\mathbf{\alpha .\hat{r}/}c\mathbf{+}\beta \tau
_{0}\}/\hslash }
\end{equation}%
where $\varepsilon $\ is real and has the dimensions of energy.

For infinitesimal transformations ($\delta \varepsilon <<1$), one can write:%
\begin{equation}
\hat{U}_{T}(\varepsilon )\simeq e^{-i(\delta \varepsilon )\{\mathbf{\alpha .%
\hat{r}}/c\}/\hbar }e^{-i(\delta \varepsilon )\beta \tau _{0}/\hbar
}=e^{-i(\delta \varepsilon )\beta \tau _{0}/\hbar }e^{-i(\delta \varepsilon
)\{\mathbf{\alpha .\hat{r}}/c\}/\hbar }
\end{equation}%
as $\ [i(\delta \varepsilon )(\mathbf{\alpha .\hat{r}}/c\hbar ),i(\delta
\varepsilon )\beta \tau _{0}/\hbar ]\propto (\delta \varepsilon )^{2}\approx
0$ (Glauber proof\cite[p.442]{Messiah}). Then the transformed Hamiltonian
can be approximated as: 
\begin{equation}
\tilde{H}_{D}=\hat{U}\hat{H}_{D}\hat{U}^{^{\dagger }}\simeq e^{i(\delta
\varepsilon )\beta m_{0}c^{2}/\hslash }e^{i(\delta \varepsilon )\mathbf{%
\alpha .\hat{r}}/c\hslash }\hat{H}_{D}e^{-i(\delta \varepsilon )\mathbf{%
\alpha .\hat{r}}/c\hslash }e^{-i(\delta \varepsilon )\beta
m_{0}c^{2}/\hslash }
\end{equation}%
Consider first:

\begin{eqnarray}
\tilde{h}_{D} &\dot{=}&e^{i(\delta \varepsilon )\mathbf{\alpha .\hat{r}}%
/c\hslash }\hat{H}_{D}e^{-i(\delta \varepsilon )\mathbf{\alpha .\hat{r}}%
/c\hslash }\simeq  \nonumber \\
&\simeq &\{I+i(\delta \varepsilon )\mathbf{\alpha .\hat{r}}/c\hslash +..\}%
\hat{H}_{D}\{I-i(\delta \varepsilon )\mathbf{\alpha .\hat{r}}/c\hslash +..\}
\nonumber \\
&\simeq &\hat{H}_{D}+i\{(\delta \varepsilon )/c\hslash \}[\mathbf{\alpha .%
\hat{r},}\hat{H}_{D}]+....
\end{eqnarray}%
Then using\cite{Thaller}:%
\begin{equation}
\lbrack \mathbf{\alpha .\hat{r},}\hat{H}_{D}\mathbf{]}\mathbf{=}3ic\hslash 
\mathit{I}\mathbf{+}2\hat{H}_{D}\{\mathbf{\alpha }-c\mathbf{\hat{p}}/\hat{H}%
_{D}\}.\mathbf{\hat{r}}
\end{equation}%
and substituting $3\mathit{I}\mathbf{=\alpha .\alpha }$, one obtains:%
\begin{eqnarray}
&&\tilde{h}_{D}\simeq \hat{H}_{D}[\mathbf{\hat{p}]}+(\delta \varepsilon )%
\mathbf{\alpha .\alpha }+i\{(\delta \varepsilon )/c\hslash \}2\{\hat{H}_{D}[%
\mathbf{\hat{p}}]\mathbf{\alpha .\hat{r}}-c\mathbf{\hat{p}}.\mathbf{\hat{r}%
\}=}  \nonumber \\
&=&c\mathbf{\alpha .\{\hat{p}+(}\delta \varepsilon )\mathbf{\alpha /}c%
\mathbf{\}+}\beta m_{0}c^{2}+i2\{(\delta \varepsilon )/c\hslash \}\{\hat{H}%
_{D}[\mathbf{\hat{p}}]\mathbf{\alpha .\hat{r}}-c\mathbf{\hat{p}}.\mathbf{%
\hat{r}\}=}  \nonumber \\
&=&\hat{H}_{D}[\mathbf{\hat{p}+(}\delta \varepsilon )\mathbf{\alpha /}c]%
\mathbf{+}i2\{(\delta \varepsilon )/c\hslash \}\{\hat{H}_{D}[\mathbf{\hat{p}}%
]\mathbf{\alpha .\hat{r}}-c\mathbf{\hat{p}}.\mathbf{\hat{r}\}}
\end{eqnarray}%
Thus, the unitary transformation induces a shift in momentum by the amount: 
\begin{equation}
\delta \mathbf{\hat{p}=\{}(\delta \varepsilon )/c\}\mathbf{\alpha =\{}%
(\delta \varepsilon )/c^{2}\}c\mathbf{\alpha }
\end{equation}%
as well as a Zitterbewegung behavior in the corresponding propagator $%
U(t)=e^{-i\tilde{H}_{D}t/\hslash }$.

For repeated infinitessimal applications one obtains a momentum displacement 
$\Delta \mathbf{\hat{p}}$ whose expectation value is%
\begin{equation}
<\Delta \mathbf{\hat{p}>}=(\varepsilon /c^{2})\mathbf{v}_{gp}=\gamma m_{0}%
\mathbf{v}_{gp}
\end{equation}%
where $\gamma =\{1-(v_{gp}/c)^{2}\}^{-1/2}$ is the Lorentz factor and $%
\mathbf{v}_{gp}$ the group velocity.

It also induces a phase shift. Indeed:%
\begin{equation}
\left\langle \Psi \right\vert \tilde{H}_{D}[\mathbf{p}]\left\vert \Psi
\right\rangle =\left\langle \Psi \right\vert e^{i(\delta \varepsilon )\beta
\tau _{0}/\hslash }\tilde{h}_{D}e^{-i(\delta \varepsilon )\beta \tau
_{0}/\hslash }\left\vert \Psi \right\rangle =\left\langle \Phi \right\vert 
\hat{H}_{D}[\mathbf{p+\alpha }\delta \varepsilon /c]\left\vert \Phi
\right\rangle +...
\end{equation}%
where%
\begin{equation}
\left\vert \Phi \right\rangle =e^{-i(\delta \varepsilon )\beta \tau
_{0}/\hslash }\left\vert \Psi \right\rangle
\end{equation}%
The phase shift is $\ \delta \varphi =(\delta \varepsilon )\beta \tau
_{0}/\hslash \ $. For a finite transformation, its expectation value is%
\begin{equation}
\left\langle \Delta \varphi \right\rangle =\{(\Delta \varepsilon )\tau
_{0}/\hslash \}\ \left\langle \beta \right\rangle =\pm (1/\gamma
)m_{0}c^{2}\tau _{0}/\hslash )
\end{equation}%
as $<\beta >=m_{0}c^{2}/<H_{D}>=\pm m_{0}c^{2}/\varepsilon =\pm 1/\gamma $ ,
for a positive (negative) energy wave packet that contains both positive and
negative energy free particle solutions\cite{Thaller}.\ Thus the sign of $%
<\beta >$ distinguishes the positive or negative energy branch where the
momentum displacement takes place. To mantain the fundamental indeterminacy
modulo $n2\pi $\ ($n$ an integer) in the phase of the complex eigenfunctions
one has to set, for $n=1$\textit{:}%
\begin{equation}
\tau _{0}=2\pi \hbar /<\beta >\varepsilon =h/m_{0}c^{2}
\end{equation}%
This is the de Broglie period. One has then:%
\begin{equation}
h/p=h/\gamma m_{0}v_{gp}=hc^{2}/\gamma m_{0}c^{2}v_{gp}=(h/\varepsilon
)(c^{2}/v_{gp})=(1/\nu )v_{ph}
\end{equation}%
which is precisely the de Broglie wave length, that is, the product of the
phase velocity by the period derived from the Planck relation $\varepsilon
=h\nu $ , as originally assumed by de Broglie\cite{Broglie}.

In conclusion, the dynamical time operator $T=\mathbf{\alpha .\hat{r}}%
/c+\beta (h/m_{0}c^{2})$\ (where the parameter $\tau _{0}$ is equated to de
Broglie period $h/m_{0}c^{2}$), generates the \textit{Lorentz boost }that
gives rise to the de Broglie wave. Indeed the fact that a rest frame
oscillation\ gives rise to a travelling wave has been shown to be a simple
consequence of special relativity applied to the complex-phase oscillation
of stationnary states through the Lorentz transformation of the time
dependence\cite{Broglie,Bayliss,Ferber}. To quote Baylis: "What in the rest
frame is a synchronized oscillation in time is seen in the laboratory to be
a wave in space".

What about Pauli's objection? The continous displacement in momentum implies
also a continous displacement in energy only within the positive and the
negative branches. No crossing of the energy gap is involved. Thus Pauli's
objection is circumvented.

Finally, it is also interesting to note the following. In the same way as
above, in the case of an infitesimal time lapse ($\delta t<<1$) \ the
unitary operator $U_{H_{D}}(\delta t)=e^{i(\delta t)\{c\mathbf{\alpha .\hat{p%
}+}\beta m_{0}c^{2}\}/\hbar }$\ can be approximated as:%
\begin{equation}
U(\delta t)\simeq e^{i(\delta t)\{c\mathbf{\alpha .\hat{p}}/\hbar
\}}e^{i(\delta t)\{\beta m_{0}c^{2}/\hbar \}}
\end{equation}%
In configuration space this yields a displacement $\delta \mathbf{r=}<%
\mathbf{\hat{r}}+c\mathbf{\alpha (}\delta t)\mathbf{>=<\hat{r}>}+\mathbf{v}%
_{gp.}(\delta t)$ and a phase shift $\delta \varphi =(\delta t)<\beta
>m_{0}c^{2}/\hbar $. For repeated infinitessimal time displacements to reach 
$\Delta t=\gamma \tau _{0}=\gamma h/m_{0}c^{2}$ (the boosted de Broglie
period), the phase shift is%
\begin{equation}
\Delta \varphi =(\gamma h/m_{0}c^{2})(m_{0}c^{2}/<H>)m_{0}c^{2}/\hbar
=h/\hbar =2\pi
\end{equation}%
These results are in agreement with the fact that the Hamiltonian is
actually the generator of the time development of a system described by a
wave packet. The approximate treatment provides only the displacement,
neglecting the dispersion of the wave.

\section{The time-energy uncertainty relation}

The time operator and the Dirac Hamiltonian satisfy the commutaion relation:%
\begin{equation}
\lbrack \hat{T},\hat{H}_{D}]=i\hbar \{I+2\beta K\}+2\beta \{\tau _{0}\hat{H}%
_{D}-m_{0}c^{2}\hat{T}\}
\end{equation}%
where $K=\beta (2\mathbf{s.l}/\hbar ^{2}+1)$\ is a constant of motion\cite%
{Thaller}. In the usual manner an uncertainty relation follows, namely:%
\begin{equation}
(\Delta T)(\Delta H_{D})\geq (\hbar /2)\left\vert \{1+2<\beta
K>\}\right\vert =(\hbar /2)\left\vert \{3+4\left\langle \mathbf{s.l}/\hbar
^{2}\right\rangle \}\right\vert 
\end{equation}%
To be noted now is that, with the present definition of the time operator,
its uncertainty is related to the uncertainty in position $\Delta \mathbf{%
\hat{r}}$ , in the same way as the energy uncertainty is related to the
momentum uncertainty $\Delta \mathbf{\hat{p}.}$ Indeed:%
\begin{eqnarray}
(\Delta T)^{2} &=&\left\langle \hat{T}^{2}\right\rangle -\left\langle \hat{T}%
\right\rangle ^{2}=\left\langle \mathbf{\hat{r}}^{2}/c^{2}+\tau
_{0}^{2}\right\rangle -\left\langle \hat{T}\right\rangle ^{2}=  \nonumber \\
&=&\{(\Delta \mathbf{r})^{2}+\left\langle \mathbf{\hat{r}}\right\rangle
^{2}\}/c^{2}+\tau _{0}^{2}-\left\langle \hat{T}\right\rangle ^{2} \\
&=&\{(\Delta \mathbf{r})^{2}+\left\langle \mathbf{\hat{r}}\right\rangle
^{2}\}/c^{2}+\tau _{0}^{2}-\left\langle \mathbf{\alpha .\hat{r}/}c\mathbf{+}%
\beta \tau _{0}\right\rangle ^{2}\geq (\Delta \mathbf{r})^{2}/c^{2} 
\nonumber
\end{eqnarray}%
\ and 
\begin{eqnarray}
\ (\Delta H_{D})^{2} &=&\left\langle \hat{H}_{D}^{2}\right\rangle
-\left\langle \hat{H}_{D}\right\rangle ^{2}= \\
&=&c^{2}\{(\Delta \mathbf{p})^{2}+\left\langle \mathbf{\hat{p}}\right\rangle
^{2}\}+(m_{0}c^{2})^{2}-\left\langle c\mathbf{\alpha .\hat{p}}+\beta
m_{0}c^{2\text{ }}\right\rangle ^{2}\geq c^{2}(\Delta \mathbf{p})^{2} 
\nonumber
\end{eqnarray}%
Then%
\begin{equation}
(\Delta \hat{T})(\Delta \hat{H})\gtrsim (\Delta \mathbf{r)(}\Delta \mathbf{%
p)\geq }(3\hbar /2)
\end{equation}

This corresponds to Bohr's interpretation: the width of a wave packet,
complementary to its momentum dispersion, measures the uncertainty in the
time of passage at a certain point, and is thereby complementary to its
energy dispersion.

\subsubsection{The Mandelstam-Tamm uncertainty relation\protect\cite[p.319]%
{Messiah}}

As an observable, the time operator can be subject to the Mandelstam-Tamm
(MT) formulation of a time-energy uncertainty relation within standard QM,
to wit: any observable $\hat{O}$ represented by a self-adjoint operator $%
\hat{O}$ not explicitly dependent on time, satisfies the dynamical equation:

\begin{equation}
(i\hbar )\frac{d}{dt}<\hat{O}>=<[\hat{O},\hat{H}]>
\end{equation}%
From the commutator\ $[\hat{O},\hat{H}]$\ it follows that the uncertainties
defined $\Delta \hat{O}$ and $\Delta \hat{H}$ satisfy the relation: 
\begin{equation}
(\Delta \hat{O})(\Delta \hat{H})\geq (1/2)\mid <[\hat{O},\hat{H}]>\mid
=(1/2)\left\vert \frac{d}{dt}<\hat{O}>\right\vert
\end{equation}%
Then, associated to any system observable $\hat{O}$ , a related time
uncertainty\textbf{\ }is\textbf{\ }defined as:%
\begin{equation}
\Delta \hat{T}_{\hat{O}}^{MT}=\frac{\Delta \hat{O}}{\mid \frac{d}{dt}<\hat{O}%
>\mid }
\end{equation}%
From Eqs.29 and 30, it then follows that:%
\begin{equation}
(\Delta \hat{T}_{\hat{O}}^{MT})(\Delta \hat{H})\geq (\hbar /2)
\end{equation}%
This is the Mandelstam-Tamm time-energy uncertainty relation. $\Delta \hat{T}%
_{\hat{O}}^{MT}$ can be interpreted as \textit{"the time required for the
center }$\left\langle \hat{O}\right\rangle $\textit{\ of this distribution
to be displaced by an amount equal to its width }$\Delta \hat{O}$\textit{"}%
\cite{Messiah}.

Now let $\hat{O}$ be the dynamical time operator in RQM $\hat{T}=(\mathbf{%
\alpha .r})/c+\beta \tau _{0}$. Then, from Eq. 32:

\begin{equation}
\Delta T_{\hat{T}}^{MT}\approx \frac{\Delta \hat{T}}{\left\vert \left\langle
I+2\beta K\right\rangle \right\vert }
\end{equation}%
It follows that:%
\begin{equation}
\frac{\Delta \hat{T}}{\left\langle I+2\beta K\right\rangle }(\Delta \hat{H}%
_{D})\geq (\hbar /2)
\end{equation}

or%
\begin{equation}
(\Delta \hat{T})(\Delta \hat{H}_{D})\geq (\hbar /2)\left\vert \left\langle
I+2\beta K\right\rangle \right\vert
\end{equation}

In the non relativistic limit $\left\langle \hat{H}_{D}\right\rangle \simeq
m_{0}c^{2}$, neglecting the oscillating terms,:

\begin{equation}
\hat{T}(t)\simeq \tau _{0}+(cp/m_{0}c^{2})^{2}t+...
\end{equation}%
Thus:%
\begin{equation}
\frac{d<\hat{T}>}{dt}=\left\langle (cp/m_{0}c^{2})^{2}\right\rangle
=(v_{gp}/c)^{2}
\end{equation}%
and%
\begin{equation}
\Delta T_{\hat{T}}^{MT}\simeq \frac{\Delta \hat{T}}{(v_{gp}/c)^{2}}\gg
\Delta \hat{T}
\end{equation}%
as $v_{gp}<<c$. The uncertainty of the Mandelstam-Tamm time operator
associated with the observable $\hat{T}$ overestimates largely the usual
uncertainty.

In the ultra relativistic limit $\left\langle \hat{H}_{D}\right\rangle
\simeq cp$: 
\begin{equation}
\hat{T}(t)\simeq t+(m_{0}c^{2}/cp)^{2}\tau _{0}+...\simeq t
\end{equation}%
and%
\begin{equation}
\Delta T_{\hat{T}}^{MT}\simeq \Delta \hat{T}
\end{equation}%
\bigskip

\section{Conclusion}

A self-adjoint internal "time operator" can be defined within Dirac's
formulation of relativistic quantum mechanics (RQM). As with all
observables, it is accorded in the Heisenberg picture a dependence on the
external laboratory time parameter $t\ $\ in the TDSE, which is atributed to
the entanglement of the microscopic system with a classical environment.\
Also as an observable, it is in general subject to an uncertainty shown to
be proportional to the position uncertainty; and consequently to a
time-energy uncertainty relation where the energy uncertainty is related to
the momentum one, supporting Bohr's original interpretation as the
uncertainty in the instant of passage at a point of the trajectory. It
nevertheless circumvents Pauli's objection, due to the fact that, as
generator of a unitary transformation, it actually produces displacements in
the continous momentum spectrum, and thus only indirectly in energy. This
resolves the problem of time in quantum mechanics.

Based on the position observable, the time operator is expected to exhibit a
Zitterbewegung behaviour about its linear dependence on $t$. As occurs with
the position Zitterbewegung, its observation is be beyond current technical
possibilities. However it may be observable in systems that simulate Dirac's
Hamiltonian, where position Zitterbewegung has allready been exhibited
experimentally\cite{Cserti,Gerritsma,LeBlanc}. A corresponding time operator
can be constructed in each case and its properties examined.

Finally, general relativity accords a dynamical behaviour to space-time,
firmly confirmed recently by the detection of gravitational waves. As a
dynamical time is definitively incompatible with a time parameter, this
becomes from the start a fundamental "problem of time" in quantum gravity%
\cite{Anderson,Isham,Butterfield}. Whether the time operator here introduced
has a relevance in this subject, is a venue to be considered\cite{Bauer3}.

\section{Appendix A. The lore of $\left[ \hat{x},\hat{p}\right] =i\hslash $}

For pedagogical purposes this Appendix collects all the properties derivable
from the commutation relation, that are usually dispersed in quantum
mechanical books.

To represent observables the operators .$\hat{x}$.and. $\hat{p}$.are
self-adjoint $(\hat{x}=\hat{x}^{\dagger }$ , $\hat{p}=\hat{p}^{\dagger })$ ,
which insures real eigenvalues. Then:

\textbf{1) Spectrum}

Consider the eigenvalue equation:

\begin{equation}
\hat{x}\ \left\vert x\right\rangle =x\ \left\vert x\right\rangle  \tag{A.1}
\end{equation}%
By Stone-von Newmann's theorem\cite{Jordan,Messiah}the operator $U(\alpha
)=\exp (-i\alpha \hat{p}/\hslash )$ with $a$ real is unitary. Then:

\begin{eqnarray*}
&&\hat{x}\ \{U(\alpha )\left\vert x\right\rangle \} \\
&=&\hat{x}\ \{1+(1/1!)(-i\alpha \hat{p}/\hslash )+(1/2!)(-i\alpha \hat{p}%
/\hslash )^{2}+(1/3!)(-i\alpha \hat{p}/\hslash )^{3}+...\}\left\vert
x\right\rangle \\
&=&(\hat{x}\ +(-i\alpha /\hslash )\hat{x}\hat{p}+(1/2)(-i\alpha /\hslash
)^{2}\hat{x}\hat{p}^{2}+...)\left\vert x\right\rangle \\
&=&(\hat{x}\ +(-i\alpha /\hslash )\{\hat{p}\hat{x}+\left[ \hat{x},\hat{p}%
\right] \}+(1/2)(-i\alpha /\hslash )^{2}\{\hat{p}^{2}\hat{x}+\left[ \hat{x},%
\hat{p}^{2}\right] \}+...)\left\vert x\right\rangle \\
&=&(\hat{x}\ +(-i\alpha /\hslash )\{\hat{p}\hat{x}+i\hslash
\}+(1/2)(-i\alpha /\hslash )^{2}\{\hat{p}^{2}\hat{x}+2i\hslash \hat{p}%
\}+...)\left\vert x\right\rangle \\
&=&\{1+(-i\alpha \hat{p}/\hslash )+(1/2)(-i\alpha \hat{p}/\hslash
)^{2}+...\}(\hat{x}+\alpha )\ \left\vert x\right\rangle \\
&=&(x+\alpha )\{1+(-i\alpha \hat{p}/\hslash )+(1/2)(-i\alpha \hat{p}/\hslash
)^{2}+...\}\left\vert x\right\rangle =(x+\alpha )\{U(\alpha )\left\vert
x\right\rangle \}
\end{eqnarray*}%
\bigskip One concludes that:%
\begin{equation}
\{U(\alpha )\left\vert x\right\rangle \}=\left\vert x+\alpha \right\rangle 
\tag{A.2}
\end{equation}

As $\alpha $\ is arbitrary, it follows that the eigenvalues of $\hat{x}\ $%
are continous from\ $-\infty $\ a $+\infty $\ , and that the eigenvectors
satisfy:%
\begin{equation}
\left\langle x^{\prime }\mid x\right\rangle =\delta (x^{\prime }-x)\ \ \ \ \
\ \ \ \ \ \ \int dx\ \left\vert x\right\rangle \left\langle x\right\vert =I 
\tag{A.3}
\end{equation}%
where $\delta (x^{\prime }-x)$\ \ \ is the Dirac delta function and $I$ is
the identity operator.

In the same way one can prove that the eigenvalues in \ $\hat{p}\ \left\vert
p\right\rangle =p\ \left\vert p\right\rangle $\ \ span a continuum from $%
-\infty $\ to $+\infty $\ and that the eigenvectors satisfy:

\begin{equation}
\left\langle p^{\prime }\mid p\right\rangle =\delta (p^{\prime }-p)\ \ \ \ \
\ \ \ \ \ \ \int dp\ \left\vert p\right\rangle \left\langle p\right\vert =I 
\tag{A.4}
\end{equation}

\textbf{2) Representations}

In configuration space one has:

\begin{equation}
\Phi (x)=\left\langle x\mid \Phi \right\rangle =\left\langle x\mid \hat{x}%
\mid \Psi \right\rangle =x\left\langle x\mid \Psi \right\rangle =x\Psi (x) 
\tag{A.5}
\end{equation}%
and:

\begin{eqnarray*}
\left\langle x^{\prime }\right\vert \left[ \hat{x},\hat{p}\right] \left\vert
x\right\rangle &=&i\hslash \delta (x^{\prime }-x)=\left\langle x^{\prime
}\right\vert \hat{x}\hat{p}-\hat{p}\hat{x}\left\vert x\right\rangle \\
&=&\int dx^{\prime \prime }\left\langle x^{\prime }\right\vert \hat{x}%
\left\vert x^{\prime \prime }\right\rangle \left\langle x^{\prime \prime
}\right\vert \hat{p}\left\vert x\right\rangle -\int dx^{\prime \prime
}\left\langle x^{\prime }\right\vert \hat{p}\left\vert x^{\prime \prime
}\right\rangle \left\langle x^{\prime \prime }\right\vert \hat{x}\left\vert
x\right\rangle \\
&=&\int dx^{\prime \prime }x^{\prime \prime }\delta (x^{\prime }-x^{\prime
\prime })\left\langle x^{\prime \prime }\right\vert \hat{p}\left\vert
x\right\rangle -\int dx^{\prime \prime }\left\langle x^{\prime }\right\vert 
\hat{p}\left\vert x^{\prime \prime }\right\rangle x\delta (x^{\prime \prime
}-x) \\
&=&x^{\prime }\left\langle x^{\prime }\right\vert \hat{p}\left\vert
x\right\rangle -x\left\langle x^{\prime }\right\vert \hat{p}\left\vert
x\right\rangle =(x^{\prime }-x)\left\langle x^{\prime }\right\vert \hat{p}%
\left\vert x\right\rangle
\end{eqnarray*}

\begin{equation}
\left\langle x^{\prime }\right\vert \hat{p}\left\vert x\right\rangle =\frac{%
i\hslash \delta (x^{\prime }-x)}{(x^{\prime }-x)}\Longrightarrow _{x^{\prime
}\rightarrow x}i\hslash \frac{d}{dx^{\prime }}\delta (x^{\prime }-x) 
\tag{A.6}
\end{equation}

Then: 
\begin{eqnarray}
\Phi (x) &=&\left\langle x\mid \Phi \right\rangle =\left\langle x\mid \hat{p}%
\mid \Psi \right\rangle =  \nonumber \\
&=&\int dx^{\prime }\left\langle x\right\vert \hat{p}\left\vert x^{\prime
}\right\rangle \left\langle x^{\prime }\mid \Psi \right\rangle =i\hslash
\int dx^{\prime }[\frac{d}{dx^{\prime }}\delta (x^{\prime }-x)]\left\langle
x^{\prime }\mid \Psi \right\rangle  \TCItag{A.7} \\
&=&\left[ \delta (x^{\prime }-x)\Psi (x^{\prime })\right] _{-\infty
}^{+\infty }-i\hslash \int dx^{\prime }\delta (x^{\prime }-x)\frac{d}{%
dx^{\prime }}\Psi (x^{\prime })=-i\hslash \frac{d}{dx}\Psi (x)  \nonumber
\end{eqnarray}%
i.e., the representation in configuration space of the vector $\left\vert
\Phi \right\rangle =\hat{p}\left\vert \Psi \right\rangle $\ is obtained by
taking the derivative of the representation of the vector $\left\vert \Psi
\right\rangle $, while the representation in configuration space of the
vector $\left\vert \Theta \right\rangle =\hat{x}\left\vert \Psi
\right\rangle $ is obtained multiplying by \ $x$\ the representation of $%
\left\vert \Psi \right\rangle $.

To conclude, in configuration space one has:%
\begin{equation}
\hat{x}\Longrightarrow x\ \ \ \ \ \ \ \ \ \ \ \ \ \hat{p}\Longrightarrow
-i\hslash \frac{d}{dx}  \tag{A.8}
\end{equation}%
and in the same way in momentum space:%
\begin{equation}
\hat{x}\Longrightarrow i\hslash \frac{d}{dp}\ \ \ \ \ \ \ \ \ \ \ \ \ \hat{p}%
\Longrightarrow \ p  \tag{A.9}
\end{equation}

\bigskip

3) \textbf{Transformation between representations}

Consider:%
\begin{equation}
\left\langle x\mid \left[ \hat{x},\hat{p}\right] \mid p\right\rangle
=i\hslash \left\langle x\mid p\right\rangle  \tag{A.10}
\end{equation}%
Developing:%
\begin{eqnarray*}
\left\langle x\mid \left[ \hat{x},\hat{p}\right] \mid p\right\rangle
&=&\left\langle x\mid \hat{x}\hat{p}\mid p\right\rangle -\left\langle x\mid 
\hat{p}\hat{x}\mid p\right\rangle \\
&=&xp\left\langle x\mid p\right\rangle -\int dx^{\prime }\left\langle x\mid 
\hat{p}\mid x^{\prime }\right\rangle \left\langle x^{\prime }\mid \hat{x}%
\mid p\right\rangle \\
&=&xp\left\langle x\mid p\right\rangle -i\hslash \int dx^{\prime }[\frac{d}{%
dx^{\prime }}\delta (x^{\prime }-x)]x^{\prime }\left\langle x^{\prime }\mid
p\right\rangle \\
&=&xp\left\langle x\mid p\right\rangle -i\hslash \lbrack \delta (x^{\prime
}-x)]x^{\prime }\left\langle x^{\prime }\mid p\right\rangle ]_{-\infty
}^{+\infty } \\
&&+i\hslash \int dx^{\prime }\delta (x^{\prime }-x)\frac{d}{dx^{\prime }}%
[x^{\prime }\left\langle x^{\prime }\mid p\right\rangle ] \\
&=&xp\left\langle x\mid p\right\rangle +i\hslash \lbrack \left\langle x\mid
p\right\rangle +i\hslash x\frac{d}{dx}\left\langle x\mid p\right\rangle ]
\end{eqnarray*}%
Substituting in Eq.(A.10) one obtains:%
\[
xp\left\langle x\mid p\right\rangle +i\hslash \lbrack \left\langle x\mid
p\right\rangle +i\hslash x\frac{d}{dx}\left\langle x\mid p\right\rangle
]=i\hslash \left\langle x\mid p\right\rangle 
\]%
\begin{equation}
i\hslash \frac{d}{dx}\left\langle x\mid p\right\rangle =-p\left\langle x\mid
p\right\rangle  \tag{A.11}
\end{equation}%
which is satisfied if:%
\begin{equation}
\left\langle x\mid p\right\rangle =Ce^{ipx/\hslash }\ \ \ \ \ \ \ \ \ \
\left\langle p\mid x\right\rangle =C^{\ast }e^{-ipx/\hslash }  \tag{A.12}
\end{equation}

Finally:%
\begin{equation}
\Phi (p)=\left\langle p\mid \Psi \right\rangle =\int dx\left\langle p\mid
x\right\rangle \left\langle x\mid \Psi \right\rangle =C^{\ast }\int dx\
e^{-ipx/\hslash }\ \Psi (x)  \tag{A.13}
\end{equation}%
and:%
\begin{equation}
\Psi (x)=\left\langle x\mid \Psi \right\rangle =\int dp\left\langle x\mid
p\right\rangle \left\langle p\mid \Psi \right\rangle =C\int dx\
e^{ipx/\hslash }\ \Phi (p)  \tag{A.14}
\end{equation}%
i.e., the representations of the state vector in the configuration and
momentum spaces are\textit{\ Fourier transforms }of each\textit{\ }other%
\textit{.} To preserve normalization one requires $C=C^{\ast }=1/\sqrt{2\pi
\hslash }$.

4) \textbf{Uncertainty relation}

Consider the state vectors%
\begin{equation}
\left\vert \Phi \right\rangle =(\hat{x}-\left\langle x\right\rangle
)\left\vert \Psi \right\rangle \ \ \ \ and\ \ \left\vert \Xi \right\rangle =(%
\hat{p}-\left\langle p\right\rangle )\left\vert \Psi \right\rangle 
\tag{A.15}
\end{equation}

Then%
\begin{equation}
\left\langle \Phi \mid \Phi \right\rangle =\left\langle \Psi \right\vert 
\hat{x}^{2}\left\vert \Psi \right\rangle -\left\langle \Psi \right\vert \hat{%
x}\left\vert \Psi \right\rangle ^{2}=(\Delta x)_{\Psi }^{2}  \tag{A.16}
\end{equation}

and%
\begin{equation}
\left\langle \Xi \mid \Xi \right\rangle =\left\langle \Psi \right\vert \hat{p%
}^{2}\left\vert \Psi \right\rangle -\left\langle \Psi \right\vert \hat{p}%
\left\vert \Psi \right\rangle ^{2}=(\Delta p)_{\Psi }^{2}  \tag{A.17}
\end{equation}

By Schawrz inequality one has%
\begin{eqnarray}
\left\langle \Phi \mid \Phi \right\rangle \left\langle \Xi \mid \Xi
\right\rangle &\geq &\left\vert \left\langle \Phi \mid \Xi \right\rangle
\right\vert ^{2}=  \nonumber \\
&=&\left\vert \left\langle \Psi \right\vert 
{\frac12}%
[\hat{x},\hat{p}]+%
{\frac12}%
\{\hat{x},\hat{p}\}-\left\langle x\right\rangle \left\langle p\right\rangle
\left\vert \Psi \right\rangle \right\vert ^{2}\geq  \nonumber \\
&\geq &\left\vert \left\langle \Psi \right\vert 
{\frac12}%
[\hat{x},\hat{p}]\left\vert \Psi \right\rangle \right\vert ^{2}=(\hslash
/2)^{2}  \TCItag{A.18}
\end{eqnarray}

Finally%
\begin{equation}
(\Delta x)_{\Psi }(\Delta p)_{\Psi }\geq \hslash /2  \tag{A.19}
\end{equation}


\begin{thebibliography}{99}
\bibitem{Pauli} Pauli, W., \textit{"The general principles of quantum
mechanics"}, Springer-Verlag, Berlin Heidelberg, footnote p.63 (1980),

\bibitem{Dirac} Dirac, P.A.M., \textit{\textquotedblleft The principles of
quantum mechanics"} (4th ed.), Oxford, Clarendon Press (1958).

\bibitem{Muga} Muga, J.G., R. Sala Mayato and I.L. Egusquiza ,
"Introduction", in J.G. Muga, R. Sala Mayato, I.L. Egusquiza (eds.) "\textit{%
Time in Quantum Mechanics"}, pp 1-28 Berlin Springer (2002); reprinted as "%
\textit{Time in Quantum Mechanics}, \textit{Vol. 1"}, Lect. Notes Phys. 
\textbf{734}, Springer-Verlag, Berlin (2008)

\bibitem{Muga2} Muga, J.G., A. Ruschhaupt and A. del Campo (eds), "\textit{%
Time in Quantum Mechanics}, \textit{Vol. 2"}, Lect. Notes Phys. \textbf{789}%
, Springer-Verlag, Berlin (2009).

\bibitem{Bauer1} Bauer, M. and P.A. Mello, "The time-energy uncertainty
relation", Ann.Phys. \textbf{111}, 38-60 (1978)

\bibitem{Bush} Busch, P., "The time-energy uncertainty relation", chapter 3
in J.G. Muga, R. Sala Mayato, I.L. Egusquiza (eds.) "\textit{Time in Quantum
Mechanics"}, Berlin Springer (2002); revised version
arXiv:quant-ph/0105049v3 (2007)

\bibitem{Briggs} Briggs, J., "A derivation of the time-energy uncertainty
relation", Journal of Physics: Conference Series \textbf{99}, 012002 (2008)

\bibitem{Hilgevoord} Hilgevoord, J., "Time in Qhantum Mrchanics", Am. J.
Phys.\textbf{\ 70}, 301-306 (2002)

\bibitem{Galapon} Galapon, E.A., "Post-Pauli's Theorem Emerging Perspective
on Time In Quantum Mechanics", Chapter 3 in\ Lect. Notes Phys. \textbf{789},
25-63 (2009)

\bibitem{Boykin} Boykin, T.B., N. Kharche and G. Klimeck, "Evolution time
and energy uncertainty", Eur. J. Phys. \textbf{28}, 673-678 (2007)

\bibitem{Bauer} Bauer, M., \textquotedblleft A dynamical time operator in
Dirac's relativistic quantum mechanics", Int.J.Mod.Phys. A \textbf{29},
1450036 (2014)

\bibitem{Briggs2} Briggs, J.S. and Jan M. Rost, \textquotedblleft Time
dependence in quantum mechanics\textquotedblright , Eur.Phys.J.\textbf{10},
(2000); \textquotedblleft On the Derivation of the Time-dependent Equation
of Schr\"{o}dinger\textquotedblright , Foundations of Physics \textbf{31},
(2001)

\bibitem{Moreva} Moreva, E. \textit{et al.}, "Time from quantum
entanglement: an experimental illustration", arXiv:1310.4691v1 [quant-ph]
(2013); "The time as an emergent property of quantum mechanics, a synthetic
description of a first experimental approach", J.Phys.: Conference Series 
\textbf{626}, 012019 (2015)

\bibitem{Thaller} Thaller, B., \textit{\textquotedblleft The Dirac Equation"}%
, Springer-Velag, Berlin Heidelberg New York (1992)

\bibitem{Greiner} Greiner, W., \textit{\textquotedblleft Relativistic
Quantum Mechanics - Wave equations"}, (3$^{\text{d}}$ ed.) Springer, Berlin
Heidelberg New York (2000)

\bibitem{Messiah} Messiah, A., \textit{"Quantum Mechanics"}, Vol.I, p. 442,
North-Holland Publishing Company, Amsterdam, and John Wiley\&Sons, New York
London Sidney, 4th printing (1966)

\bibitem{Jordan} Jordan, T.J., \textit{"Linear operators for quantum
mechanics"}, John Wiley \& Sons, Inc. (1969)

\bibitem{Broglie} de Broglie, L., Ph.D. thesis; Ann. Phys.,Ser. 10$^{\text{e}%
}$, t. III (1925). English translation reprinted in Ann.Fond.Louis de
Broglie \textbf{17}, p. 92 (1992)

\bibitem{Bayliss} Baylis, W.E., \textquotedblleft De Broglie waves as an
effect of clock desynchronization", Can.J.Phys. \textbf{85}, pp.1317-1323
(2007)

\bibitem{Ferber} Ferber, R., "A Missing Link: What is behind de Broglie's
"periodic phenomenon"?, Found.Phys.Lett. \textbf{9}, 575-586 (1996)

\bibitem{Lan} Lan, S.Y. et al., "A Clock Directly Linking Time to a Particle
Mass", Science \textbf{339}, 554-557 (2013)

\bibitem{Catillon} Catillon, P. \textit{et al.}, \textquotedblleft A Search
for the de Broglie Particle Internal Clock by means of Electron
Channeling\textquotedblright , Found.Phys. \textbf{38}, 659-664 (2008)

\bibitem{Cserti} Cserti, J. and G. David, "Unified description of
Zitterbewegung for spintronic, graphene, and superconducting systems",
Phys.Rev. B \textbf{74}, 172305 (2006)

\bibitem{Gerritsma} Gerritsma, R. \textit{et al}., \textquotedblleft Quantum
simulation of the Dirac equation", Nature \textbf{463}, pp.68-71 (2010)

\bibitem{LeBlanc} LeBlanc, L.J. \textit{et al}., \textquotedblleft Direct
observation of Zitterbewegung in a Bose-Einstein condensate", New J.Phys. 
\textbf{15}, 073011 (2013)

\bibitem{Anderson} Anderson, E., "The problem of time in quantum gravity",
Ann.Phys. (Berlin) \textbf{524}, 757-786 (2012) and references therein; also
arXiv:1009.2157v3 [gr-qc]

\bibitem{Isham} Isham, C.J., "Canonical Quantum Gravity and the Problem of
Time", arXiv:gr-qc/9210011v1, (1992); "Prima Facie Questions in Quantum
Gravity", arXiv:gr-qc/9310031v1 (1993)

\bibitem{Butterfield} Butterfield, J. and C.J. Isham, "On the Emergence of
Time in Quantum Gravity", arXiv:gr-qc/9901024v1 (1999)

\bibitem{Bauer3} Bauer, M., "Quantum Gravity and a Time Operator in
Relativistic Quantum Mechanics", arXiv:gr-qc/1605.01659 (2016)
\end{thebibliography}
\end{document}